\documentclass[conference]{IEEEtran}

\usepackage{graphicx} 
\usepackage{amsmath,amssymb,mathrsfs}
\usepackage{balance}

\ifCLASSINFOpdf

\else

\fi

\begin{document}
\title{SA-MIMO: Scalable Quantum-Based \\Wireless Communications}

\author{Jiuyu Liu, Yi Ma$^\dagger$, and Rahim Tafazolli\\{Institute for Communication Systems (ICS), 5G/6G Innovation Centre} \\{University of Surrey, Guildford, U.K.}\\{E-mail: (jiuyu.liu, y.ma, r.tafazolli)@surrey.ac.uk}
		}
{}

\maketitle

\begin{abstract}	
Rydberg atomic receivers offer a quantum-native alternative to conventional RF front-ends by directly detecting electromagnetic fields via highly excited atomic states. 
While their quantum-limited sensitivity and hardware simplicity make them promising for future wireless systems, extending their use to scalable multi-antenna and multi-carrier configurations, termed Scalable Atomic-MIMO (SA-MIMO), remains largely unexplored. 
This paper introduces a novel RF transmitter–atomic receiver architecture that addresses this gap. 
The core idea lies in a novel modulation technique called Phase-Rotated Symbol Spreading (PRSS), which transforms the nonlinear phase retrieval problem inherent to atomic detection into a tractable linear demultiplexing task. 
PRSS enables efficient signal processing and supports scalable MUX/DeMUX operations in both atomic MIMO and atomic OFDM systems. 
Simulation results show that the proposed system achieves up to 2.5 dB gain under optimal maximum-likelihood detection and over 10 dB under suboptimal detection in MIMO settings. 
These results establish PRSS-assisted SA-MIMO as a promising architecture for realizing high-sensitivity, interference-resilient atomic wireless communication.
\end{abstract}

\begin{IEEEkeywords}
Atomic receiver, quantum wireless, scalability, multi-antenna (MIMO), multi-carrier transmission. 
\end{IEEEkeywords}

\IEEEpeerreviewmaketitle

\section{Introduction}
\begin{figure}[t!]
    \centering
    \includegraphics[width=0.49\textwidth]{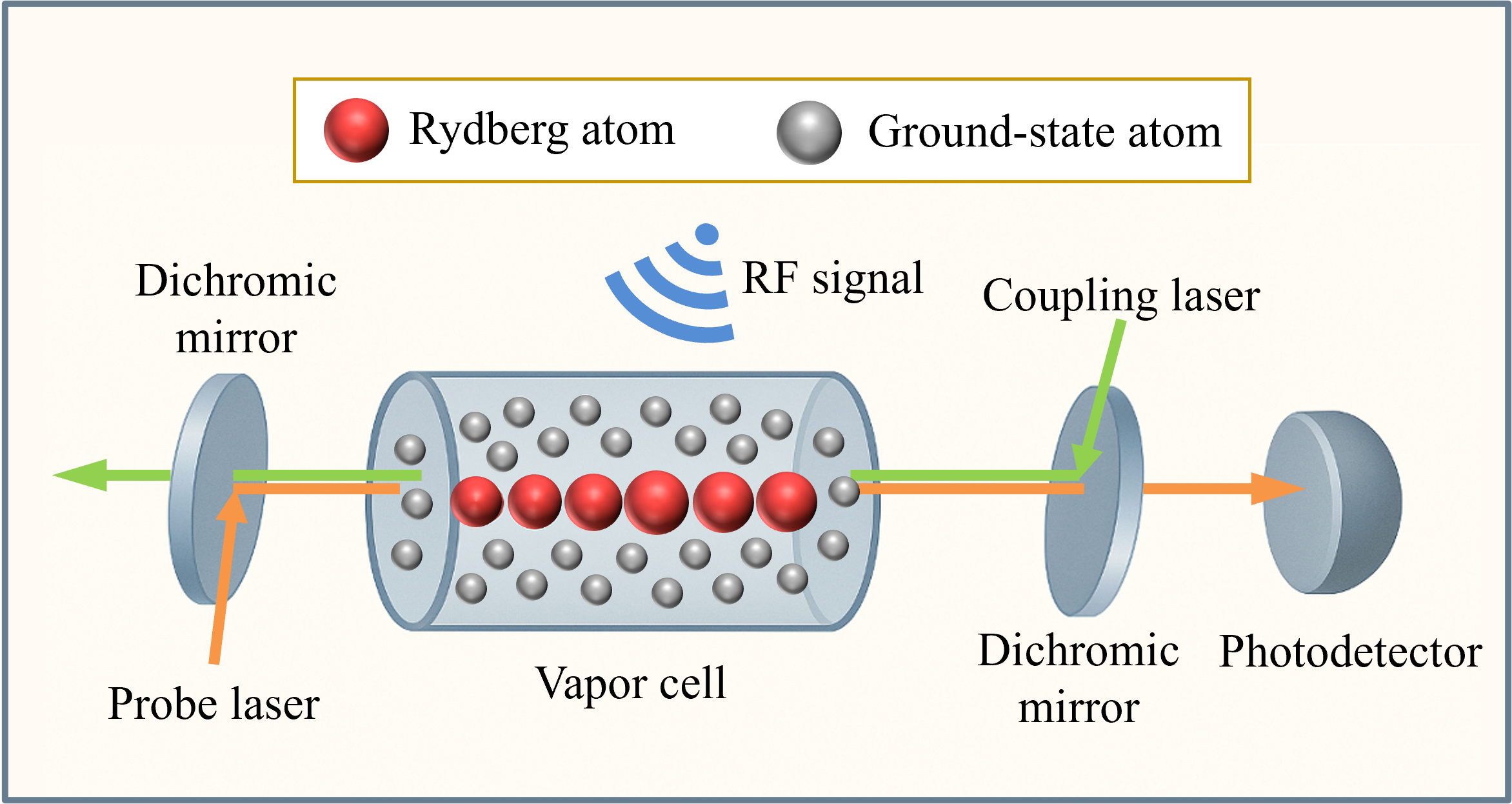} 
    \caption{Schematic of a Rydberg atomic receiver based on electromagnetically induced transparency (EIT).}\label{receiver}
    \vspace{-1em}
\end{figure}

Rydberg atomic receivers (illustrated in Fig. \ref{receiver}) represent a fundamentally novel approach to radio-frequency (RF) signal detection. 
These receivers leverage quantum sensing principles to overcome intrinsic limitations associated with conventional semiconductor-based front-end architectures. 
Traditional RF front-ends typically depend on low-noise amplifiers, mixers, and filters, which inherently introduce thermal noise and nonlinear distortion. 
In contrast, atomic receivers detect electromagnetic fields directly through highly excited atomic states known as Rydberg atoms.

This direct quantum-level interaction enables a set of unique capabilities, including quantum-limited sensitivity, extremely narrow spectral selectivity, and broad frequency tunability \cite{Fancher2021}. 
These advantages are achieved without relying on local oscillators or frequency downconversion circuitry. 
As a result, Rydberg atomic receivers are considered promising candidates for next-generation wireless communication systems, especially in scenarios that demand high energy efficiency, enhanced sensitivity to weak signals, or strong resilience to interference \cite{Meyer2018}. 
Their integration has the potential to unlock ultra-low-power wireless links, extend coverage in challenging environments, enable precise spectrum access, and support novel physical-layer designs that move beyond the limitations of classical RF electronics.

Despite these appealing advantages, the practical integration of atomic receivers into wireless systems remains a major challenge. 
Most existing studies have focused on single-antenna, single-carrier configurations that do not meet the spatial and spectral efficiency requirements of modern communication networks (e.g., \cite{Meyer2018,Xu2025, Meyer2023,Anderson2021}). 
Extending atomic reception to multi-antenna systems, referred to as Atomic-MIMO (A-MIMO) in \cite{Cui2024, Cui2024a, Cui2025}, and to multi-carrier (MC) systems introduces several new technical hurdles. 
These include signal multiplexing and demultiplexing (MUX/DeMUX), demodulation, channel estimation, and the efficient handling of computational complexity. 
Additionally, the inherent sensitivity of atomic receivers must be carefully managed in the presence of multi-dimensional interference and realistic wireless impairments.

This paper introduces a novel RF transmitter–atomic receiver architecture that supports scalable MUX/DeMUX operations in A-MIMO systems, laying the foundation for what we term Scalable Atomic-MIMO (SA-MIMO). 
The paper begins with a review of the operating principles of atomic receivers, followed by their extension to A-MIMO and an analysis of the nonlinear phase retrieval problem encountered during signal detection. 
The scalability limitations of current A-MIMO implementations are then examined from a signal processing perspective.

To overcome these limitations, we propose a novel RF modulation technique termed phase-rotated symbol spreading (PRSS). 
In PRSS, each modulation symbol is spread across two time slots with a deterministic phase offset. 
This structured phase relationship enables the inherently nonlinear phase retrieval problem to be reformulated as a linear demultiplexing task in the de-spreading domain, which can be efficiently solved using conventional linear techniques.
With careful design of the phase offset, the proposed system outperforms existing A-MIMO solutions, achieving a performance gain of up to $2.5$~dB under MLD and over $10$~dB under suboptimal detection approaches.
Encouraged by these results, we further extend the PRSS approach to atomic receiver-based multi-carrier systems. The resulting architecture demonstrates a performance gain of up to $20$~dB compared to conventional RF-based OFDM systems under comparable conditions.

\section{Fundamentals and Challenges in Atomic Receiver Systems}
This section introduces the fundamental principles behind Rydberg atomic receivers and their extension to A-MIMO systems. 
It also outlines the scalability challenges that arise in multi-antenna and multi-carrier scenarios, motivating the architectural and algorithmic innovations proposed in this work.
 
\subsection{Principles of Rydberg Atomic Receivers}
Rydberg atomic receivers detect RF signals by leveraging the extreme sensitivity of highly excited Rydberg states to external electromagnetic fields. 
The detection mechanism relies on the phenomenon of electromagnetically induced transparency (EIT) in a four-level atomic system. 
A weak probe laser and a strong coupling laser are used to establish quantum interference that renders the medium transparent at the probe wavelength \cite{Zhang2023d}.

As illustrated in Fig. \ref{receiver}, a typical atomic receiver setup includes a vapor cell containing alkali atoms (e.g., rubidium or cesium), where the probe laser is tuned near a ground-to-intermediate optical transition, and the coupling laser excites the atom further into a Rydberg state.
When these lasers satisfy the EIT condition, the probe laser experiences minimal absorption, resulting in a narrow transparency window in the optical transmission spectrum. 
This effect is highly sensitive to perturbations introduced by RF fields that couple nearby Rydberg states.

When an RF electric field $E_{\mathrm{RF}}(t)$ is applied and is resonant with a transition between two Rydberg levels, 
it induces Autler–Townes splitting (see \cite{autler1955stark}) in the EIT resonance. 
The resulting transmission spectrum splits into two peaks, separated by a frequency gap $\Delta\omega$, which is directly proportional to the RF field amplitude:
\begin{equation}\label{eq01}
\Delta\omega=\frac{\mu E_{\mathrm{RF}}}{h_P},
\end{equation}
where $\mu$ is the electric dipole moment for the Rydberg-Rydberg transition, and $h_P$ is the reduced Planck constant.

The atomic receiver does not directly output voltages or currents like a classical antenna. 
Instead, it converts the RF signal into a change in optical transparency, which is detected using a photodetector. 
The photodetector measures the intensity of the probe laser after it passes through the vapor cell, generating a voltage or current proportional to the transmitted optical power. 
By sweeping the probe laser frequency and recording the detector output, the EIT transmission spectrum is obtained. 
In the absence of RF, a single peak is observed. 
When an RF field is present, two peaks appear, and their frequency separation $\Delta\omega$ reveals the RF field strength.
Thus, the Rydberg atomic receiver indirectly senses the RF signal via a nonlinear optical transformation of the field amplitude. 
The final output is typically related to the square of the RF signal envelope \cite{Zhang2023d}:
\begin{equation}\label{eq02}
y(t)\propto|E_{\mathrm{RF}}(t)|^2.
\end{equation}

\subsection{Atomic-MIMO Signal Model}
Consider an A-MIMO system with $N$ RF transmit antennas, each transmitting a complex baseband signals $x_n(t), n=1, ..., N$. 
The resulting RF field at an atomic receiver is given by:
\begin{equation}\label{eq03}
E_{\mathrm{RF}}(t)=\Re\left\{\sum_{n=1}^Nh_nx_n(t)e^{j2\pi f_c t}\right\},
\end{equation}
where $h_n\in\mathbb{C}$ represents the complex channel gain from the $n$-th transmit antenna, and $f_c$ is carrier frequency. 

The atomic receiver interacts with the total incident RF field through a nonlinear optical process, resulting in an output that depends on the envelope of the field. Under the assumption of slow-varying $x_n(t)$ relative to the RF carrier, and considering the aggregate effect of RF noise, atomic decoherence, and quantum shot noise, the received signal can be modeled as \cite{Cui2025}:
\begin{equation}\label{eq04}
	y(t)=\left|\sum_{n=1}^Nh_nx_n(t)+v(t)\right|,
\end{equation}
where $v(t)\sim CN(0,\sigma^2)$ represents additive white Gaussian noise as justified by invoking the law-of-large-numbers \cite{Cui2025}.

For notational simplicity, we omit the time index $t$ in \eqref{eq04} and introduce a receiver index $k$. 
Assuming there are $K$ atomic receivers, we represent the received signal at the $k$-th receiver as:
\begin{equation}\label{eq05}
y_k = \left| \mathbf{h}_k^T \mathbf{x} + v_k \right|,~ k=1,..., K,
\end{equation}
where $\mathbf{h}_k = [h_{k,1}, \dots, h_{k,N}]^T$, $\mathbf{x} = [x_1, \dots, x_N]^T$, and $[\cdot]^T$ denotes matrix/vector transpose.

Stacking the signals from all $K$ receivers, we form the received vector:
$\mathbf{y} = [y_1, \dots, y_K]^T$.
Given full knowledge of $\mathbf{h}_k$ for all $k$, the A-MIMO signal detection problem is to estimate the transmitted signal vector $\mathbf{x}$ as a function of observed signals and channel vectors:
\begin{equation}\label{eq06}
\hat{\mathbf{x}} = f(\mathbf{y}, \{\mathbf{h}_k\}_{k=1}^K).
\end{equation}

\subsection{Signal Detection Algorithms and Scalability Analysis}\label{2c}
Given the nonlinear magnitude-based model in \eqref{eq05}, the signal detection task is nontrivial due to the loss of phase information and the non-convexity of the absolute-value operation. 
In this section, we briefly review both optimal and low-complexity signal detection algorithms, and discuss their scalability with the size of A-MIMO.

\subsubsection{Maximum Likelihood Detection (MLD)}
Assuming the transmitted signal vector $\mathbf{x}$ belongs to a known discrete constellation set $\mathcal{X} \subset \mathbb{C}^N$ (e.g., QAM or PSK), the MLD algorithm seeks the signal vector that minimizes the squared error between the observed envelope and the predicted envelope \cite{kay1998detection}:
\begin{equation}\label{eq07}
\hat{\mathbf{x}}_{\mathrm{MLD}} = \underset{\mathbf{x} \in \mathcal{X}}{\arg\min} \sum_{k=1}^{K} \left( y_k - \left| \mathbf{h}_k^T \mathbf{x} \right| \right)^2.
\end{equation}
This approach is optimal in the maximum-likelihood sense; 
however, it suffers from exponential complexity in $N$ due to exhaustive search over $\mathcal{X}$, making it impractical for large-scale A-MIMO systems or high-order modulations.

\subsubsection{Envelope Matched Filtering (EMF)}
Analogous to the matched filtering algorithm commonly used in linear systems, EMF can be interpreted as a heuristic matched filter in the nonlinear (envelope) domain. Specifically, each received envelope $y_k$ is treated as a scaled projection of the transmitted signal vector $\mathbf{x}$ along the direction of $\mathbf{h}_k$, leading to the approximation:
\begin{equation}\label{eq08}
\tilde{\mathbf{x}} = \sum_{k=1}^K y_k \cdot \frac{\mathbf{h}_k}{\|\mathbf{h}_k\|},
\end{equation}
where $\|\cdot\|$ denotes the Euclidean norm. 
A hard decision is then made via symbol-wise quantization:
\begin{equation}\label{eq09}
\hat{x}_n = \underset{s \in \mathcal{S}}{\arg\min} \left| \tilde{x}_n - s \right|, \quad n=1,\dots,N,
\end{equation}
where $\mathcal{S}$ is the scalar modulation constellation.

This algorithm avoids nonlinear optimization, relying only on linear weighting and nearest-neighbor mapping. Its computational complexity is $\mathcal{O}(NK)$, i.e., linear in both the number of transmit antennas and atomic receivers. 
EMF is particularly effective when the channel vectors $\{\mathbf{h}_k\}$ are approximately orthogonal and the noise level is moderate \cite{kay1998detection}.

However, the EMF approach is inherently sub-optimal. It discards phase information and assumes that the observed envelopes faithfully represent the projection magnitudes of $\mathbf{x}$, without accounting for channel correlations or interference. 
As a result, its performance can degrade significantly in interference-limited environments or when the channel vectors are highly correlated.

\subsubsection{Envelope Squared Least Squares (ES-LS)}
ES-LS plays as an intermediate approach between EMF and MLD. 
We consider squaring both sides of \eqref{eq05}:
\begin{equation}\label{eq10}
y_k^2 = \left| \mathbf{h}_k^T \mathbf{x} + v_k \right|^2 \approx \left| \mathbf{h}_k^T \mathbf{x} \right|^2 + w_k,
\end{equation}
where $w_k$ models the residual noise and cross-terms. This yields a nonlinear least-squares problem:
\begin{equation}\label{eq11}
\hat{\mathbf{x}}_{\mathrm{ES-LS}} = \arg\min_{\mathbf{x} \in \mathcal{X}} \sum_{k=1}^K \left( y_k^2 - \left| \mathbf{h}_k^T \mathbf{x} \right|^2 \right)^2.
\end{equation}
This forms a nonlinear and non-convex least squares problem due to the modulus-squared terms $\left| \mathbf{h}_k^T \mathbf{x} \right|^2$. 
The structure is similar to the magnitude-squared regression problem that appears in phase retrieval and $1$-bit quantized MIMO systems \cite{9540292}.

To enable tractable optimization, we may relax the constraint $\mathbf{x} \in \mathcal{X}$ and allow $\mathbf{x} \in \mathbb{C}^N$.
This relaxation allows standard iterative process \cite{Cook12018}:
\begin{equation}\label{eq12}
\mathbf{x}^{(i+1)} = \mathbf{x}^{(i)} - \eta \sum_{k=1}^K \left( \left| \mathbf{h}_k^T \mathbf{x}^{(i)} \right|^2 - y_k^2 \right) \cdot \mathbf{h}_k \mathbf{h}_k^H \mathbf{x}^{(i)},
\end{equation}
where $\eta > 0$ is a step size.
Once the relaxed solution $\hat{\mathbf{x}}$ is obtained, each component can be mapped onto the nearest constellation symbol.

The ES-LS method achieves a trade-off between the computational simplicity of EMF and the accuracy of MLD. 
When optimized with gradient descent, the complexity scales as $\mathcal{O}(NKI)$, where $I$ is the number of iterations.

\subsubsection{Expectation-Maximization with Gibbs Sampling (EM-GS)}
This algorithm models the unobserved complex-valued projections at the receiver as latent variables, and iteratively estimates the transmitted signal vector $\mathbf{x}$ by maximizing the expected complete-data likelihood \cite{casella1992explaining}.

Let $z_k = \mathbf{h}_k^T \mathbf{x} + v_k$ denote the latent complex-valued observation prior to envelope detection. The EM-GS algorithm proceeds by alternating between the following two steps:

\paragraph*{E-Step (Gibbs Sampling)}
In the E-step, the algorithm samples from the posterior distribution $p(\mathbf{z}|\mathbf{y}, \mathbf{x}^{(i)})$, where $\mathbf{z} = [z_1, ..., z_K]^T$ are the latent complex observations and $\mathbf{x}^{(i)}$ is the current estimate. 
Since the conditional distribution is analytically intractable, Gibbs sampling is employed to generate samples from the complex-valued $z_k$ conditioned on the magnitude constraint $|z_k| = y_k$ and the Gaussian noise model.

\paragraph*{M-Step (Signal Update)}
In the M-step, a new estimate $\mathbf{x}^{(i+1)}$ is computed by maximizing the expected complete-data log-likelihood:
\begin{equation}\label{eq13}
\mathbf{x}^{(i+1)} = \underset{\mathbf{x} \in \mathcal{X}}{\arg\min} \sum_{k=1}^K \mathbb{E}_{z_k} \left[ \left| z_k - \mathbf{h}_k^T \mathbf{x} \right|^2 \right],
\end{equation}
where the expectation is approximated using the Gibbs samples. This step reduces to a least-squares optimization over the signal space, which can be solved in closed form or approximated via projection onto the discrete constellation $\mathcal{X}$.

The EM-GS algorithm is particularly powerful in handling the nonlinear detection problem under strong noise and channel correlation. 
By explicitly modeling the unobserved phase through probabilistic inference, EM-GS achieves performance close to MLD, especially when the number of Gibbs samples per iteration is sufficiently large.

However, this improvement comes at the cost of increased computational complexity. Let $L$ denote the number of Gibbs samples per iteration and $I$ the number of EM iterations. 
The overall complexity scales as $\mathcal{O}(LINK)$, which is not suitable for large A-MIMO systems.

\begin{figure}[t!]
    \centering
    \includegraphics[width=0.31\textwidth]{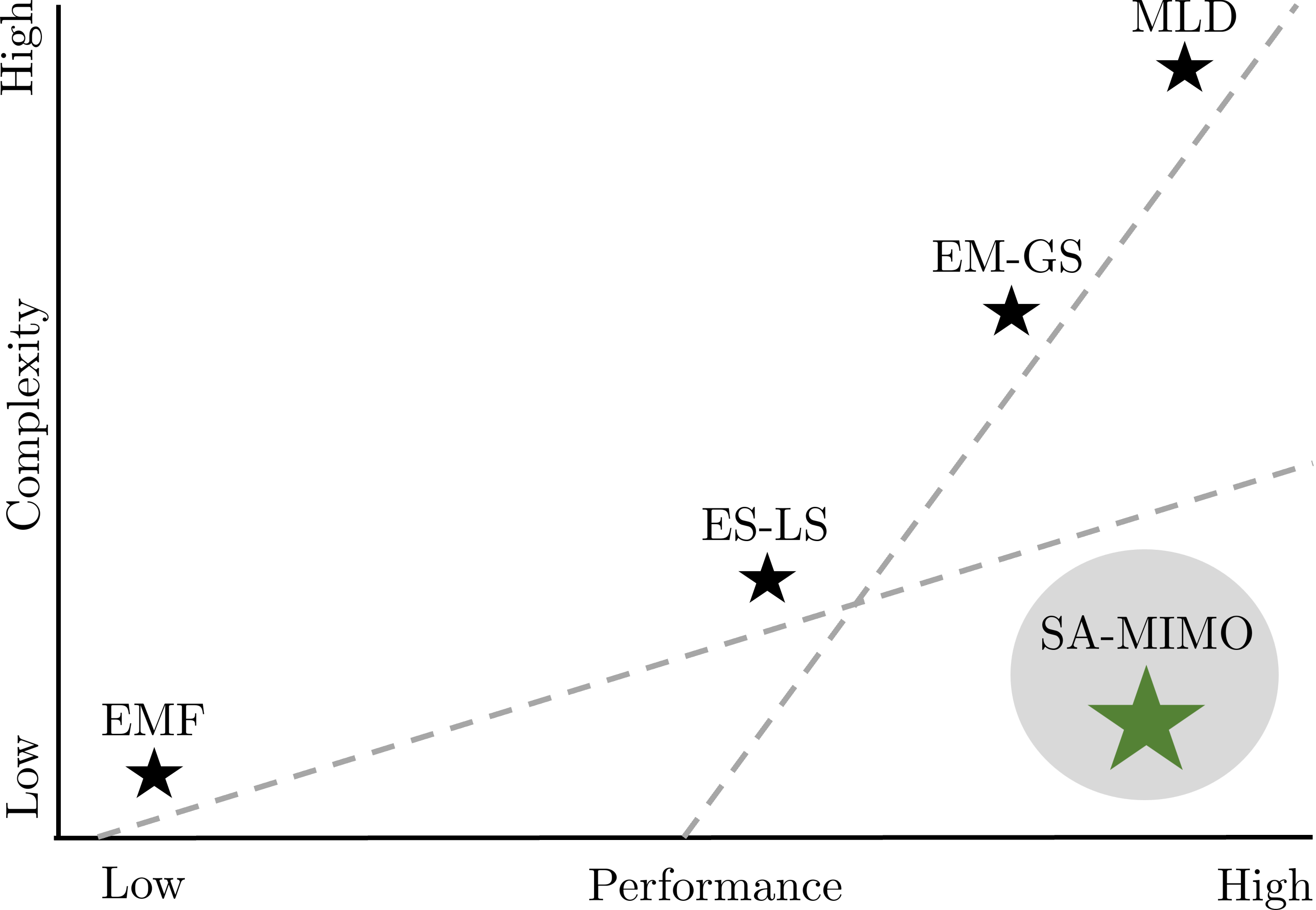} 
    \caption{Performance-complexity trade-off of A-MIMO detection algorithms.}\label{tradeoff}
        \vspace{-10pt}
\end{figure}

In summary, Fig.~\ref{tradeoff} illustrates the performance–complexity trade-off of the A-MIMO detection algorithms discussed in this section. 
Unfortunately, none of the existing methods achieves the desirable balance between performance and computational complexity, as indicated by the shaded grey region in the plot. 
This observation motivates the proposed SA-MIMO architecture, which will be introduced in Section~\ref{sec3}.

\section{The SA-MIMO Architecture}\label{sec3}
In this section, we begin with the reference signal injection approach introduced in \cite{Cui2025}, upon which we build our proposed SA-MIMO architecture.

\subsection{Reference Injection}
The idea of reference injection originates from holographic MIMO systems, where a known reference signal is spatially embedded into the radiated field to enable robust noncoherent detection. 
In the context of A-MIMO systems, where receivers are limited to envelope detection, the loss of phase information severely hampers coherent signal recovery. 
Reference injection alleviates this challenge by superimposing a known complex-valued bias term onto the received field, effectively acting as a local oscillator.

Mathematically, the received signal at the $k$-th atomic receiver with reference injection $r_k$ is modeled as:
\begin{equation}\label{eq:ref_model}
y_k = |\mathbf{h}_k^T \mathbf{x} + v_k+ r_k |.
\end{equation}
The injected reference $r_k$ can be spatially modulated across $k$ to improve diversity and identifiability.
Since the phase of $r_k$ can be absorbed into $\mathbf{h}_k$ and $v_k$ without loss of generality, we assume $r_k \in \mathbb{R}^+$ throughout the remainder of this work.

This formulation shifts the nonlinear detection problem into a biased phase retrieval problem. 
Compared to the original envelope-only model, the reference-injected model offers two key advantages:

\subsubsection{Improved Identifiability} The known complex bias $r_k$ reduces the ambiguity in the envelope measurement, enhancing the observability of $\mathbf{x}$, especially under channel correlation or limited receiver diversity.
\subsubsection{Algorithmic Convergence} Detection algorithms introduced in Section \ref{2c} benefit from more structured observations, resulting in better convergence behavior and estimation accuracy.

While reference injection enhances detection robustness, it remains a receiver-centric solution. 
To further improve the flexibility, scalability, and adaptability of A-MIMO systems under practical constraints, we propose a novel architecture, SA-MIMO, which introduces the concept of phase-rotated symbol spreading at the transmitter side.

\subsection{PRSS-Assisted SA-MIMO}
\begin{figure}[t!]
    \centering
    \includegraphics[width=0.48\textwidth]{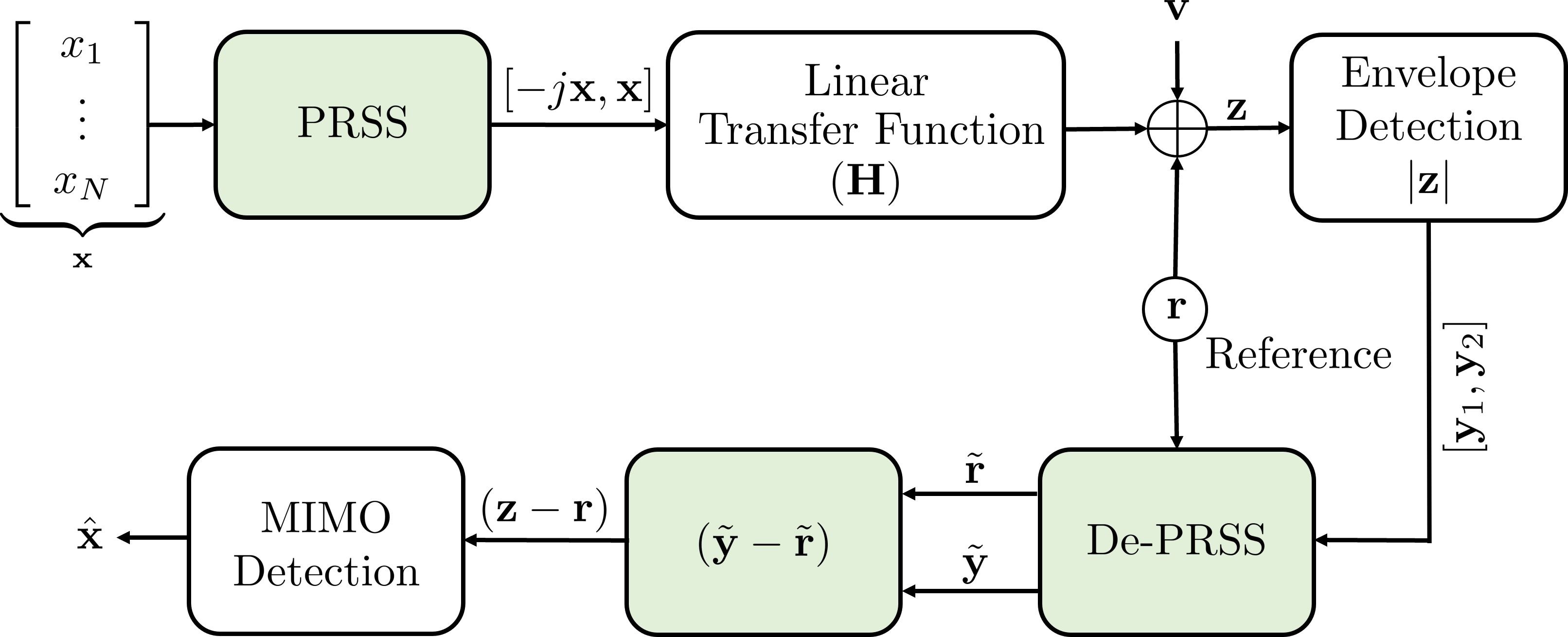} 
    \caption{Schematic of the PRSS-assisted SA-MIMO architecture. This architecture is suitable for both multiuser MIMO and point-to-point MIMO.}
    \label{amimo}
    \vspace{-7pt}
\end{figure}

Fig.~\ref{amimo} illustrates the proposed PRSS-assisted SA-MIMO architecture. 
The core idea is to apply time-domain symbol spreading at each transmitter, followed by independent de-spreading at each atomic receiver. We will show that, through this mechanism, the nonlinear biased phase retrieval problem is effectively transformed into a linear multiplexing problem.

\subsubsection{PRSS-Assisted Transmission}
Let $\mathbf{c} = [1, \; e^{j\theta}]^T$, where $\theta \neq 0$, denote the spreading code employed identically by all transmitters. This code is used to generate the time-domain spread signal:
\begin{equation}\label{eq15}
\tilde{\mathbf{x}}_n = x_n \mathbf{c},
\end{equation}
The resulting vector $\tilde{\mathbf{x}}_n$ is transmitted over two consecutive time slots: in the first slot, $x_n$ is transmitted, and in the second slot, $x_n e^{j\theta}$ is transmitted.

\subsubsection{PRSS De-spreading}
Assuming that the wireless channel remains constant across these two time slots, 
the received signals at the $k$-th atomic receiver over two consecutive slots can be expressed as:
\begin{IEEEeqnarray}{ll}
y_{k,1}&=|\underbrace{\mathbf{h}_k^T \mathbf{x} + v_{k,1}}_{=z_{k,1}}+ r_k |\label{eq16},\\
y_{k,2}&=|\underbrace{\mathbf{h}_k^T \mathbf{x}e^{j\phi} + v_{k, 2}}_{=z_{k,2}}+ r_k |.\label{eq17}
\end{IEEEeqnarray}

We let $\phi=\frac{3\pi}{2}$ and represent \eqref{eq16}-\eqref{eq17} into
\begin{IEEEeqnarray}{ll}
y_{k,1}&=|(\Re(z_{k,1})+r_k)+j\Im(z_{k,1}) |\label{eq18},\\
y_{k,2}&=|(\Im(z_{k,2})+r_k)-j\Re(z_{k,2}) |\label{eq19}.
\end{IEEEeqnarray}
In A-MIMO systems, the injected reference  $r_k$ is typically generated in close physical proximity to the atomic receiver, and thus satisfies the condition: 
\begin{equation}\label{eq20}
r_k \gg |z_{k,1}|~ \mathrm{and}~ r_k \gg |z_{k,2}|.
\end{equation}
Under this sufficient condition, we can drop the corresponding imaginary terms in \eqref{eq18}-\eqref{eq19} and obtain:
\begin{IEEEeqnarray}{ll}
y_{k,1} &\approx r_k + \Re\{z_{k,1}\}=r_k+\Re(\mathbf{h}_k^T\mathbf{x})+\Re(v_{k,1}), \label{eq21}\\
y_{k,2} &\approx r_k + \Im\{z_{k,2}\}=r_k+\Im(\mathbf{h}_k^T\mathbf{x})+\Im(v_{k,2}). \label{eq22}
\end{IEEEeqnarray}
Combining \eqref{eq21} and \eqref{eq22} using the operation \( y_{k,1} + j y_{k,2} \), we obtain:
\begin{equation}\label{eq23}
\mathbf{h}_k^T\mathbf{x}+v_k
=\underbrace{(y_{k,1}+jy_{k,2})}_{=\tilde{y}_k}-\underbrace{(r_k+jr_k)}_{=\tilde{r}_k}, ~_{k=1,...,K}
\end{equation}
where \( v_k = \Re\{v_{k,1}\} + j \Im\{v_{k,2}\} \) is the effective complex-valued noise term after de-spreading.

Finally, by stacking all \( K \) receivers, we express the PRSS-assisted SA-MIMO system in the following vector form:
\begin{equation}\label{eq24}
\tilde{\mathbf{y}} - \tilde{\mathbf{r}} = \mathbf{H} \mathbf{x} + \mathbf{v},
\end{equation}
where $\tilde{\mathbf{y}}=[\tilde{y}_1,..., \tilde{y}_K]^T$, $\tilde{\mathbf{r}}=[\tilde{r}_1,..., \tilde{r}_K]^T$, $\mathbf{v}=[v_1, ...,v_K]^T$, and $\mathbf{H}=[\mathbf{h}_1, ...,\mathbf{h}_K]^T$.
This exhibits a standard linear system model, where conventional signal detection algorithms such as MLD, zero-forcing (ZF), linear minimum mean square error (LMMSE), and matched filtering (MF) can be directly applied.

\subsection{Performance Analysis and Extension}\label{3c}
\subsubsection{Capacity of PRSS-Assisted SA-MIMO}
In this section, we compare the envelope-only A-MIMO and the proposed PRSS-assisted SA-MIMO architectures in terms of their channel capacity and computational complexity.

The channel capacity for the PRSS-assisted SA-MIMO architecture is straightforward to compute. 
Due to the use of PRSS, each symbol is transmitted over two time slots, effectively halving the spectral efficiency. 
As a result, the achievable capacity is given by
\begin{equation}\label{eq25}
C_{\text{prss}} = \frac{1}{2} \log_2 \det \left( \mathbf{I}_K + \frac{G_{\mathrm{atom}}}{\sigma_{\mathrm{RF}}^2} \mathbf{H} \mathbf{Q} \mathbf{H}^H \right),
\end{equation}
where $\mathbf{Q} = \mathbb{E}[ \mathbf{x} \mathbf{x}^H ]$ denotes the input covariance matrix.
To highlight the advantage of atomic receivers, we introduce the parameter $\sigma_{\mathrm{RF}}^2$, denoting the noise power of conventional RF receivers, and the parameter $G_{\mathrm{atom}}$, capturing the SNR advantage of atomic receivers over their RF counterparts. 

Experimental results reported in recent physics literature (e.g., \cite{Meyer2018}) suggest that atomic receivers can achieve an SNR gain $G_{\mathrm{atom}}$ often exceed $20$ dB for some certain operational bandwidths.
As a result, the capacity $C_{\text{PRSS}}$ can be substantially higher than that of conventional RF-based MIMO systems, especially in the low-to-moderate SNR regime.

\subsubsection{Capacity of Envelop-Only A-MIMO}
While no general closed-form capacity (or mutual information) expression exists for the envelope-only model due to the nonlinear modulus operation, a tight high-SNR approximation can be constructed by summing the scalar mutual information of independent envelope-detected links. Specifically, the scalar channel
\begin{equation}\label{eq26}
y = |x + v|, \quad x \sim \mathcal{CN}(0, \mathcal{E}_x), \quad v \sim \mathcal{CN}(0, \sigma^2)
\end{equation}
has the mutual information:
\begin{equation}\label{eq27}
C_{\mathrm{env}} = \hbar(y) - \hbar(y|x),
\end{equation}
with $\hbar(y)$ denoting the differential entropy of the output amplitude, and $\hbar(y|x)$ the conditional entropy given the input.

The output distribution $p_y(y)$ follows a Rayleigh distribution:
\begin{equation}\label{eq28}
p_y(y) = \frac{2y}{\mathcal{E}_x + \sigma^2}\exp\left(-\frac{y^2}{\mathcal{E}_x+\sigma^2}\right), \quad y \geq 0,
\end{equation}
from which the differential entropy is numerically approximated as \cite{nielsen2010entropies}:
\begin{equation}\label{eq29}
\hbar(y) = \log_2(\sqrt{\mathcal{E}_x+\sigma^2})+1+\frac{\gamma}{2}
\end{equation}
where \( \gamma\approx0.5772 \) is the Euler–Mascheroni constant.

Analogously, we can obtain the conditional entropy $\hbar(y|x)$:
\begin{equation}\label{eq30}
\hbar(y|x)=\hbar(|v|)=\log_2(\sqrt{\sigma^2})+1+\frac{\gamma}{2}
\end{equation}
Substituting \eqref{eq29}-\eqref{eq30} into \eqref{eq27}, we obtain:
\begin{equation}\label{eq31}
C_{\mathrm{env}}=\frac{1}{2}\log_2\left(1+\frac{\mathcal{E}_x}{\sigma^2}\right).
\end{equation}

This result represents the mutual information for a scalar envelope channel with Gaussian input. While this input is not capacity-achieving for the nonlinear channel, it provides a tight lower bound and enables meaningful comparisons. If the MIMO channel matrix $\mathbf{H}$ is unitary and the input $\mathbf{x} \sim \mathcal{CN}(0, \mathcal{E}_x\mathbf{I})$, the system decouples into $K$ independent scalar channels of the form \eqref{eq26}, and the total mutual information becomes:
\begin{equation}\label{eq32}
C_{\mathrm{env}}^{\text{MIMO}} \approx \frac{K}{2} \log_2\left(1 + \frac{\mathcal{E}_x}{\sigma^2}\right).
\end{equation}

This expression shares the same functional form as the capacity of the PRSS-assisted SA-MIMO system in \eqref{eq25}. This implies that, under the Gaussian-input assumption and an idealized unitary channel structure, both systems exhibit similar mutual information scaling, despite their differing physical architectures and receiver implementations.
Therefore, even with time-domain symbol spreading, the PRSS-assisted approach does not compromise spectral efficiency in theory. This conclusion will be further supported by our simulation results in Section~\ref{sec4}.

\subsubsection{Complexity and Extendability}
With the linear model in \eqref{eq24} established, the receiver-side processing for PRSS-assisted SA-MIMO aligns closely with that of conventional MIMO systems. 
The only additional computational step is the de-spreading operation, which involves a simple vector subtraction
$(\tilde{\mathbf{y}} - \tilde{\mathbf{r}})$. 
This operation incurs negligible complexity overhead compared to the overall detection pipeline.

It is also worth noting that the spreading and de-spreading operations are independent of the structure of the channel multiplexing matrix $\mathbf{H}$. 
For example, consider the case where $\mathbf{H} = \mathbf{C} \mathbf{F}^H$, where $\mathbf{C}$ is a Toeplitz matrix commonly used to model frequency-selective channels in OFDM systems, and $\mathbf{F}^H$ is the normalized inverse discrete Fourier transform (IDFT) matrix (with \( [\cdot]^H \) denoting Hermitian transpose). 
Under this structure, the system model directly reduces to that of a standard OFDM receiver, where efficient FFT-based processing can be immediately applied for signal demultiplexing. The proposed PRSS-assisted architecture, therefore, integrates naturally with multi-carrier systems without modifying the underlying FFT-based detection chain.

In addition, the spreading and de-spreading operations do not require any channel state information. 
This provides a further advantage in terms of channel estimation, as all conventional estimation techniques can be directly applied to the linear model in \eqref{eq24}.

\begin{figure}[t!]
    \centering
    \includegraphics[width=0.48\textwidth]{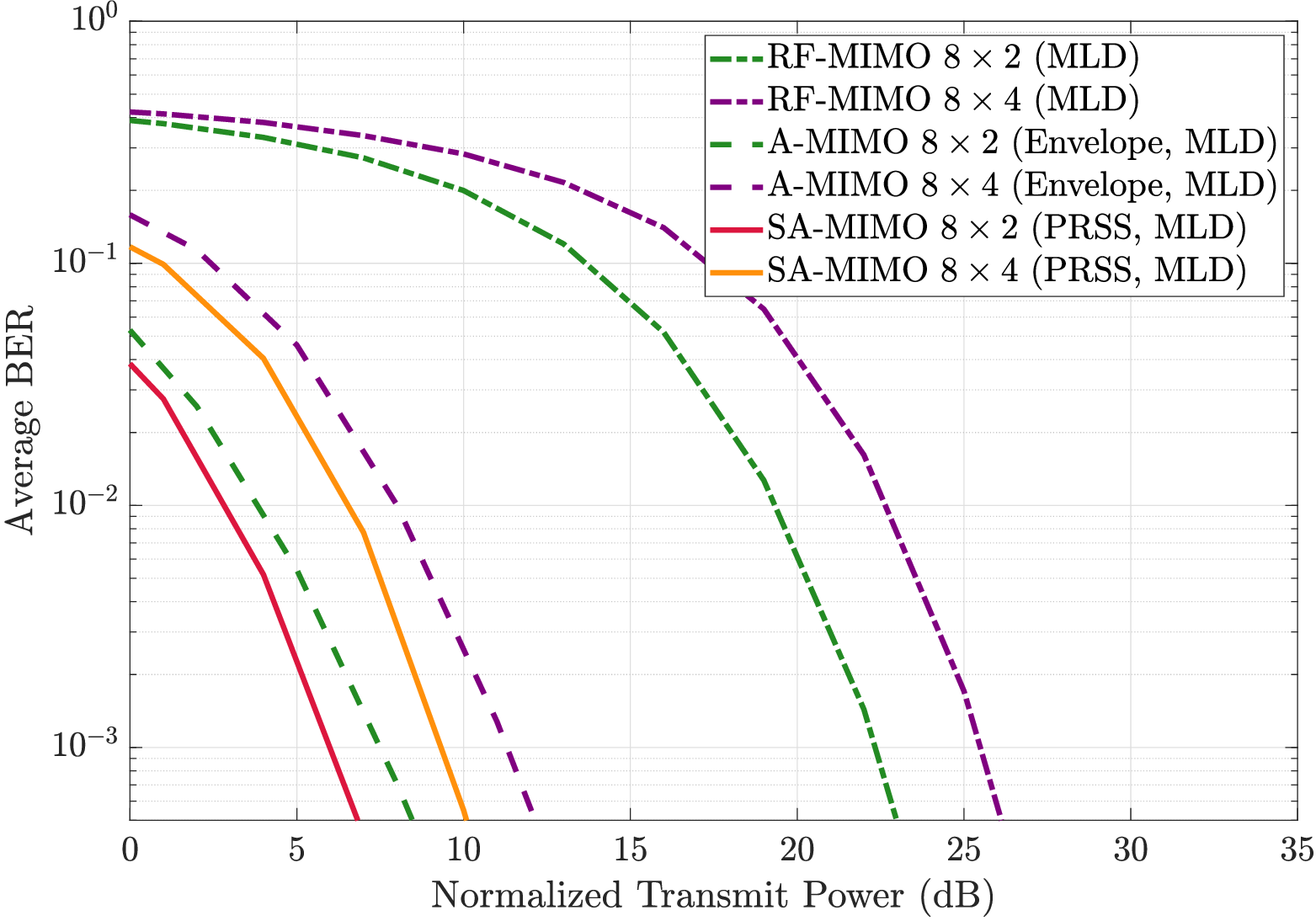} 
    \caption{Average BER vs transmit power normalized by noise for small-scale MIMO configurations ($N=2, 4$ and $M=8$) using optimum MLD algorithms.}
    \label{fig1}
    \vspace{-10pt}
\end{figure}

\section{Simulation Results and Discussion}\label{sec4}
This section presents simulation results that validate and extend the theoretical analysis of the proposed PRSS-assisted SA-MIMO architecture. 
We focus on uncoded MIMO and OFDM systems employing 4-QAM and 16-QAM modulation schemes, with bit error rate (BER) as the primary performance metric. 
The injected reference signal is $35$ dB higher than the communication signal. 
Each data point is obtained through extensive Monte Carlo simulations to ensure statistical reliability. 
The study comprises three experiments, each designed to assess different performance aspects under varying algorithmic assumptions and system scales.

\subsubsection*{Experiment 1 (Optimal MLD Performance)}

This experiment evaluates the BER performance of the PRSS-assisted SA-MIMO system using the optimal MLD algorithm. 
As a benchmark, we compare against the envelope-only A-MIMO system, also employing MLD.
The wireless channel is modeled as an i.i.d. Rayleigh fading MIMO channel with complex Gaussian coefficients. 
To ensure fair comparison in spectral efficiency, the PRSS-assisted system employs 16-QAM (transmitting one symbol over two time slots), while the envelope-only system uses 4-QAM.

The results in Fig.~\ref{fig1} show that the BER performance depends on the number of transmit antennas ($N$), exhibiting the well-known trade-off between the spatial diversity and multiplexing gain, which is consistent with conventional MIMO systems.
For a fixed MIMO configuration, the PRSS-assisted SA-MIMO outperforms the envelope-only A-MIMO by approximately 2.5~dB, confirming our theoretical claim in Section~\ref{3c} that PRSS preserves spectral efficiency while significantly enhancing robustness.

For further reference, we also include the performance of an RF-MIMO system under the same MIMO configuration, with a receiver SNR gain $G_{\mathrm{atom}} = 20 $~dB based on reported atomic receiver measurements \cite{Meyer2018}. 
The considerable gap between RF-MIMO and SA-MIMO illustrates the potential SNR or coverage gain enabled by atomic receiver technologies, particularly in power-limited or long-range scenarios.

\subsubsection*{Experiment 2 (Suboptimal Detection Performance)}

This experiment compares the performance of PRSS-assisted and envelope-only A-MIMO under suboptimal detection schemes.
For both the PRSS-assisted and conventional RF MIMO systems, we employ the projected Jacobi (p-Jacobi) detector proposed in \cite{Liu2023}, which achieves near-MLD performance with only square-order computational complexity.
In contrast, the envelope-only baseline uses the EM-GS algorithm, which delivers the best BER performance among suboptimal methods for envelope-based detection, albeit at a significantly higher computational cost (see Fig.~\ref{tradeoff}).

\begin{figure}[t!]
    \centering
    \includegraphics[width=0.48\textwidth]{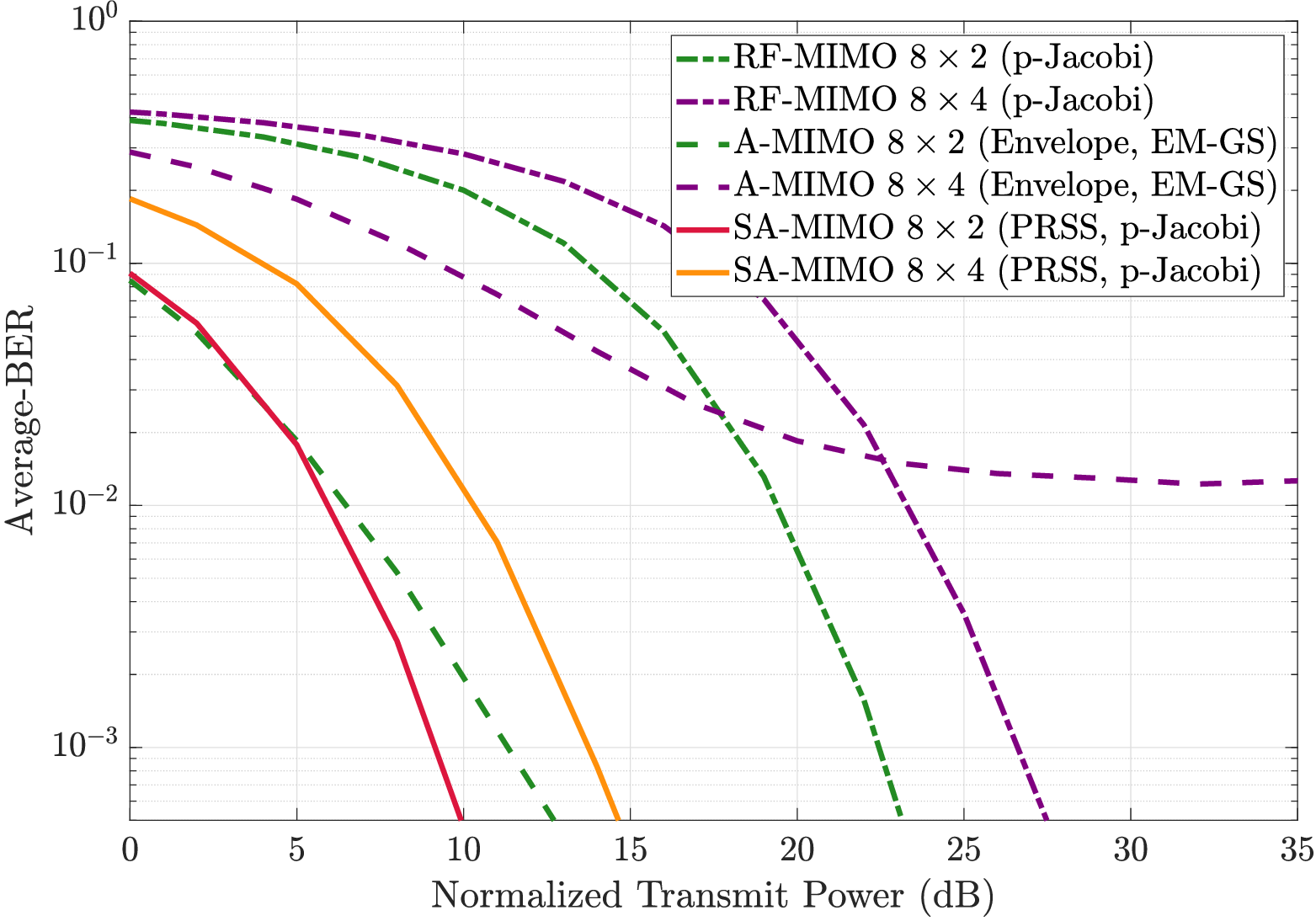} 
    \caption{Average BER vs normalized transmit power for small-scale MIMO configurations (\( N=2, 4 \); \( M=8 \)) using suboptimal detection algorithms.}
    \label{fig2}
    \vspace{-2pt}
\end{figure}

As shown in Fig.~\ref{fig2}, for $N=2$, PRSS and envelope-only models exhibit similar performance at low SNRs ($< $5~dB). However, the PRSS-assisted system yields up to 3~dB gain at higher SNRs. 
When the number of transmit antennas increases to $N=4$, the envelope-only A-MIMO performance degrades significantly, even exhibiting an error floor beyond 20~dB. 
In contrast, the PRSS-assisted model maintains a steady BER slope, resulting in up to 10~dB performance gain.

Due to the exponential complexity of MLD, {\it Experiment 1} was limited to small-scale MIMO settings. 
In this experiment, we also explore the scalability of PRSS-assisted SA-MIMO under large-scale configurations using sub-optimum detectors. 
We simulate configurations with $N=48, 64$ transmit antennas and $M=128$ atomic receivers.

\begin{figure}[t!]
    \centering
    \includegraphics[width=0.48\textwidth]{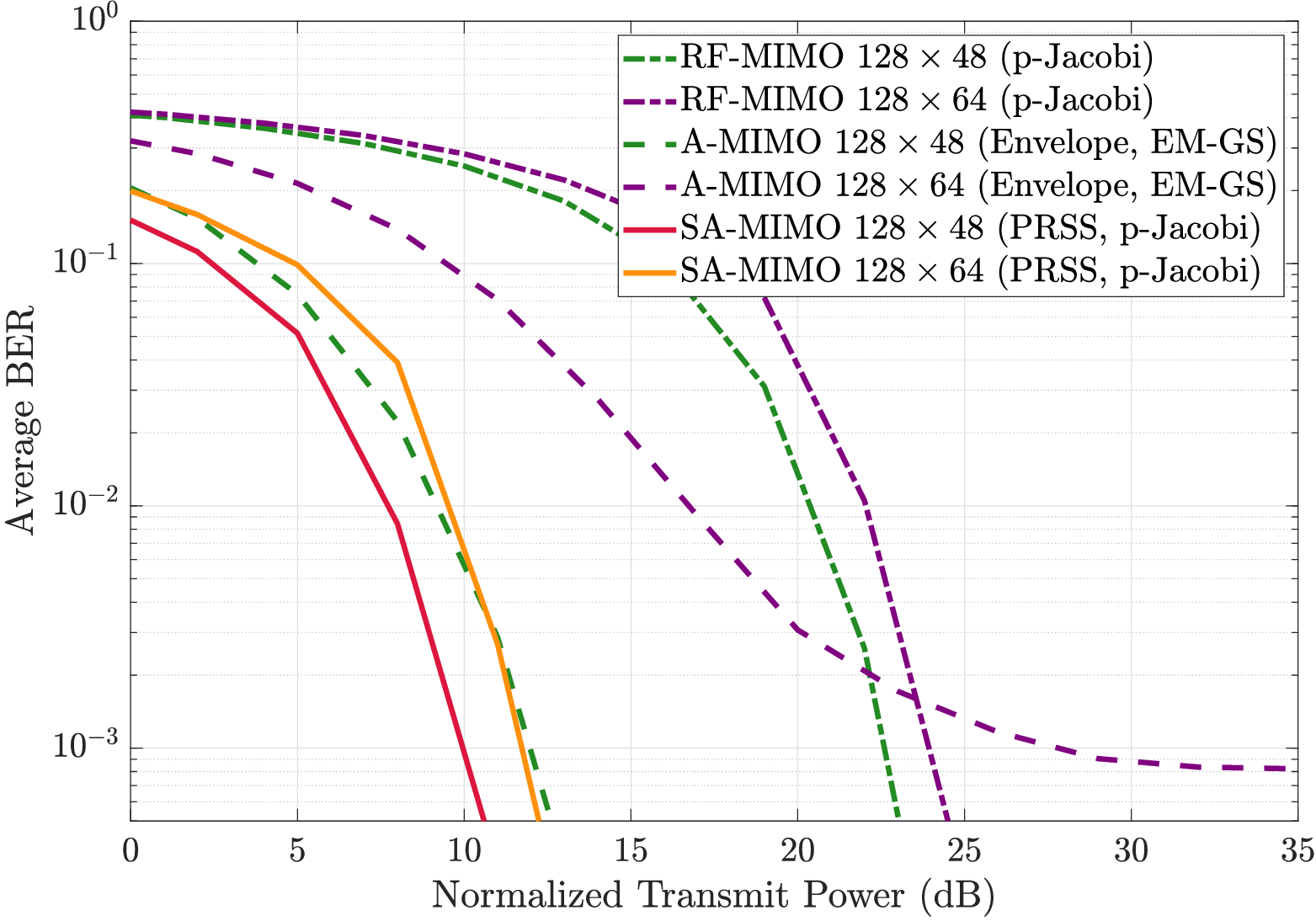} 
    \caption{Average BER vs normalized transmit power for large-scale MIMO configurations (\( N=48, 64 \); \( M=128 \)), using suboptimal algorithms.}
    \label{fig3}
    \vspace{-0pt}
\end{figure}

Fig.~\ref{fig3} clearly demonstrates that the PRSS-assisted system consistently outperforms the envelope-only system in all large-scale settings. 
As the number of transmit antennas increases, the envelope-only system suffers severe performance degradation and error floors, while PRSS scales gracefully. In the case of $N=64$, the performance gap reaches 20~dB, underscoring the superior scalability and efficiency of the proposed architecture.

\subsubsection*{Experiment 3 (PRSS Assisted Atomic OFDM Receiver)}
This experiment investigates the performance of a PRSS-assisted atomic OFDM receiver. 
We consider a point-to-point communication system where a single antenna RF transmitter sends a PRSS-assisted OFDM signal using 64 subcarriers. 
Three modulation schemes are evaluated: 4QAM, 16QAM, and 256QAM. 
The system is assumed to be uncoded in order to isolate the modulation and detection performance. 
This setup allows us to clearly assess the behavior of the DFT based receiver under the PRSS architecture.
We do not include an envelope-only system as a baseline in this experiment, since such systems are incompatible with DFT receivers due to their nonlinear nature.

\begin{figure}[t!]
    \centering
    \includegraphics[width=0.48\textwidth]{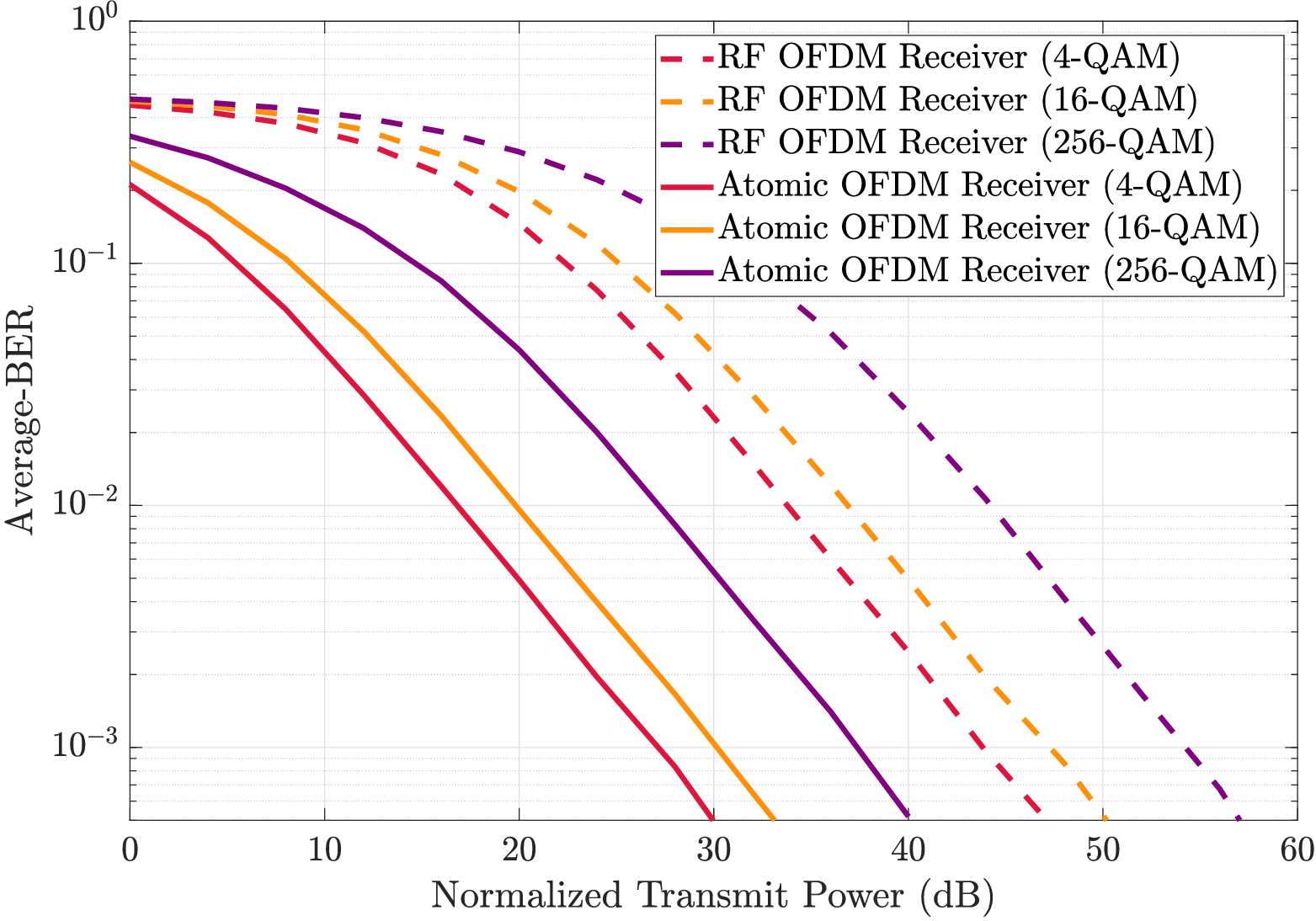} 
    \caption{Average BER vs normalized transmit power for PRSS-assisted Atomic OFDM receivers (No. of Subcarriers = 64 ) under Rayleigh fading channel.}
    \label{fig4}
    \vspace{-0pt}
\end{figure}

As shown in Fig.~\ref{fig4}, the PRSS architecture supports DFT processing effectively. 
The BER performance of the atomic system closely follows that of the conventional RF based OFDM system. 
The main performance difference arises from the higher effective SNR offered by the atomic receiver.

\section{Conclusion}
This paper presented a novel PRSS-assisted SA-MIMO architecture that enables scalable and efficient signal detection for atomic receiver based wireless MIMO systems. 
Through both theoretical study and computer simulations, the PRSS-assisted architecture demonstrates clear advantages over envelope-only A-MIMO systems in both bit error performance and complexity. 
It preserves spectral efficiency, supports standard linear detection methods, and remains robust in large antenna configurations. These attributes make it a practical and high performance candidate for future wireless systems based on atomic receivers.

\ifCLASSOPTIONcaptionsoff
  \newpage
\fi

\bibliographystyle{IEEEtran}
\bibliography{../QuantumMIMO}

\end{document}